\documentclass[aps,prb,twocolumn,amsmath,amssymb]{revtex4}

\usepackage{array}
\usepackage{amssymb}
\usepackage{amsmath}
\usepackage{color}
\usepackage{bm}
\usepackage{graphics}
\usepackage{setspace}
\newskip\humongous 
\def\caja{\mathsurround=0pt}
\def\eqalign#1{\,\vcenter{\openup1\jot \caja	
\ialign{\strut \hfil$\displaystyle{##}$&$	
\displaystyle{{}##}$\hfil\crcr#1\crcr}}\,}
\newif\ifdtup

\def\ref#1{$^{#1}$}

\begin{document}
\title{Efficient stochastic thermostatting of path integral molecular dynamics}

\author{Michele Ceriotti}
\email{michele.ceriotti@gmail.com}
\affiliation{Computational Science, Department of Chemistry and Applied Biosciences, ETH Z\"urich, USI Campus, Via Giuseppe Buffi 13, CH-6900 Lugano, Switzerland}

\author{Michele Parrinello}
\affiliation{Computational Science, Department of Chemistry and Applied Biosciences, ETH Z\"urich, USI Campus, Via Giuseppe Buffi 13, CH-6900 Lugano, Switzerland}

\author{Thomas E. Markland and David E. Manolopoulos}
\affiliation{Physical and Theoretical Chemistry Laboratory, Oxford
University, South Parks Road, Oxford OX1 3QZ, UK}

\date{\today}

\begin{abstract}
The path integral molecular dynamics (PIMD) method provides a convenient way to compute the quantum mechanical structural and thermodynamic properties of condensed phase systems at the expense of introducing an additional set of high-frequency normal modes on top of the physical vibrations of the system. Efficiently sampling such a wide range of frequencies provides a considerable thermostatting challenge. Here we introduce a simple stochastic path integral Langevin equation (PILE) thermostat which exploits an analytic knowledge of the free path integral normal mode frequencies.  We also apply a recently-developed colored-noise thermostat based on a generalized Langevin equation (GLE), which automatically achieves a similar, frequency-optimized sampling. The sampling efficiencies of these thermostats are compared with that of the more conventional Nos\'e-Hoover chain (NHC) thermostat for a number of physically relevant properties of the liquid water and hydrogen-in-palladium systems. In nearly every case, the new PILE thermostat is found to perform just as well as the NHC thermostat while allowing for a computationally more efficient implementation. The GLE thermostat also proves to be very robust delivering a near-optimum sampling efficiency in all of the cases considered. We suspect that these simple stochastic thermostats will therefore find useful application in many future PIMD simulations.
\end{abstract}

\maketitle

\section{Introduction}

Atomistic computer simulations are now routinely used to shed light on the behavior of condensed phase systems of interest in physics, chemistry, biology and materials science. Typically these simulations assume that the nuclei can be treated as classical particles, even when the interactions between them are obtained using {\em ab initio} methods. For systems comprised of heavy atoms at high temperatures, this is often quite a reasonable assumption. However, whenever any light atoms such as hydrogen are present at room temperature or below, quantum mechanical zero point energy and tunneling effects in the nuclear motion can play an important role in determining the properties of the system.

Since the 1980s it has been possible to include quantum mechanical effects in simulations of structural and thermodynamic properties such as radial distribution functions and average potential and kinetic energies by performing imaginary time path integral\cite{Feynman65,Feynman72} (PI) simulations. These simulations exploit the isomorphism between the quantum mechanical partition function of the physical system and the classical partition function of an extended problem consisting of a necklace of replicas of the system connected by harmonic springs.\cite{Chandler81} 

Early implementations of this approach used Monte Carlo (MC) methods to sample the configuration space,\cite{Barker79} but these methods often proved to be inefficient unless carefully constructed collective moves were used. It was then suggested by Parrinello and
Rahman\cite{Parrinello82,Parrinello84} that momenta could be assigned to the replicas, allowing the sampling to be performed by molecular dynamics (MD). In principle, this allows many of the techniques originally developed for use in classical MD simulations to be used in a path integral context. 

In practice, the stiff harmonic springs between the replicas lead to inefficient and nonergodic dynamics when microcanonical trajectories are used to generate ensemble averages.\cite{Hall84} A simple stochastic Andersen thermostatting scheme\cite{Andersen80} was originally suggested to overcome this problem.\cite{Berne86} However, stochastic approaches have since largely been abandoned following the development of deterministic Nos{\'e}-Hoover chain thermostats,\cite{Nose84,Hoover85,Martyna92} which have now become the gold standard for performing path integral molecular dynamics (PIMD) simulations.\cite{Tuckerman93} These thermostats generate ergodic, canonical averages and provide a conserved quantity which can be used to check the integration time step, at the expense of introducing sets of auxiliary chain variables which add to the complexity of the calculation.\cite{Martyna92,Tuckerman93}

In the last few years stochastic approaches have regained attention in a variety of different contexts, and many of their drawbacks have been overcome. For example, a strategy to define a conserved quantity has been developed which not only allows one to check the quality of the integration but also allows for the correction of sampling errors due to the use of a finite time step.\cite{Bussi07b} Indeed the total energy of the system minus the accumulated heat absorbed from the thermostat will clearly be a conserved quantity for any thermostatting scheme. Moreover new stochastic methods have been introduced which show very promising sampling properties.\cite{Bussi07a,Ceriotti09a} In particular, a stochastic velocity rescaling thermostat has been developed which couples to the total kinetic energy of the system rather than the individual momenta of the particles within it.\cite{Bussi07a} This allows canonical sampling to be achieved with a much smaller disturbance to the Hamiltonian trajectory. In addition, it has been shown how a generalized (colored-noise) Langevin equation thermostat can be tuned to sample a very wide range of frequencies simultaneously and efficiently.\cite{Ceriotti09a,Ceriotti10}

The purpose of the present paper is to exploit these recent developments by introducing two new stochastic path integral thermostats that are competitive in terms of sampling efficiency with the Nos{\'e}-Hoover chain thermostat but simpler to implement and cheaper to use. To demonstrate this, we have performed an extensive set of benchmark calculations, computing the statistical sampling efficiencies of various thermostats for a number of physical observables in two very different condensed phase systems: a flexible model of liquid water at room temperature and the diffusion of an interstitial hydrogen atom in palladium.

The outline of the paper is as follows. Section II begins by reviewing the theory of PIMD and discussing how the equations of motion for the path integral replicas (or ring polymer beads) can be integrated in the absence of a thermostat. This sets the scene for the remainder of the section, which introduces two new stochastic thermostats for path integral simulations, compares them with the well-established Nos\'e-Hoover chain thermostat, and ends with a brief discussion about the difference between global and local thermostatting. Section III presents the results of our calculations on the liquid water and hydrogen-in-palladium systems and Section~IV presents our conclusions.

\section{Theory}

\subsection{Path integral molecular dynamics}

Consider a system of distinguishable particles described by a Cartesian Hamiltonian of the form
$$
H = \sum_{i=1}^{N} {p_i^2\over 2m_i}+V(q_1,\ldots,q_{N}), \eqno(1)
$$
in which the potential energy $V(q_1,\ldots,q_{N})$ is such that the quantum mechanical partition function 
$$
Z = {\rm tr}\left[e^{-\beta H}\right]
\eqno(2)
$$
is well defined (with $\beta=1/k_{\rm B}T$). After a standard Trotter product\cite{Trotter59,Schulman05} discretization of the trace, this partition function can be written as
$$
Z \simeq {1\over (2\pi\hbar)^{f}}\int d^f{\bf p}\,\int d^f{\bf q}\,e^{-\beta_nH_n({\bf p},{\bf q})},
\eqno(3)
$$
where $f=Nn$ and $\beta_n=\beta/n$. Here $H_n({\bf p},{\bf q})$ is the classical Hamiltonian of a fictitious ring polymer consisting of $n$ copies of the system connected by harmonic springs,\cite{Feynman72,Chandler81}
$$
H_n({\bf p},{\bf q}) = H_n^0({\bf p},{\bf q})+V_n({\bf q}), \eqno(4)
$$
where
$$
H_n^0({\bf p},{\bf q}) = \sum_{i=1}^{N}\sum_{j=1}^n \left({[{p}_{i}^{(j)}]^2\over 2m_i}+{1\over 2}m_i\omega_n^2[q_{i}^{(j)}-q_{i}^{(j-1)}]^2\right), \eqno(5)
$$
and
$$
V_n({\bf q}) = \sum_{j=1}^n V({q}_{1}^{(j)},\ldots,{q}_{N}^{(j)}), \eqno(6)
$$
with $\omega_n=1/\beta_n\hbar$ and $q_{i}^{(0)}\equiv q_{i}^{(n)}$. The error in Eq.~(3) is $O(1/n^2)$ and so vanishes in the limit as $n\to\infty$.

The path integral molecular dynamics (PIMD) method\cite{Parrinello84} uses this classical isomorphism as a computational tool to calculate quantum mechanical equilibrium properties of the form
$$
\left<A\right> = {1\over Z}{\rm tr}\left[e^{-\beta H}A\right]. \eqno(7)
$$
For example, the potential energy of the system is given by
$$
\left<V\right> \simeq {1\over (2\pi\hbar)^fZ}\int d^f{\bf p}\int d^f{\bf q}\,
e^{-\beta_nH_n({\bf p},{\bf q})}{\cal V}_n({\bf q}), \eqno(8)
$$
where the estimator ${\cal V}_n({\bf q})$ involves an average over the beads of the ring polymer necklace,
$$
{\cal V}_n({\bf q}) = {1\over n}V_n({\bf q}) = {1\over n}\sum_{j=1}^n V({q}_{1}^{(j)},\ldots,{q}_{N}^{(j)}), \eqno(9)
$$
and the kinetic energy can be calculated as 
$$
\left<T\right> \simeq {1\over (2\pi\hbar)^fZ}\int d^f{\bf p}\int d^f{\bf q}\,
e^{-\beta_nH_n({\bf p},{\bf q})}{\cal T}_n({\bf q}), \eqno(10)
$$
where ${\cal T}_n({\bf q})$ is most efficiently chosen to be the centroid virial estimator,\cite{Herman82,Ceperley95}
$$
{\cal T}_n({\bf q}) = {N\over 2\beta}+{1\over 2n}\sum_{i=1}^{N}\sum_{j=1}^n ({q}_{i}^{(j)}-\overline{q}_i)\cdot {\partial V({q}_{1}^{(j)},\ldots,{q}_{N}^{(j)})\over\partial {q}_{i}^{(j)}}, \eqno(11)
$$
with
$$
\overline{q}_i = {1\over n}\sum_{j=1}^n {q}_{i}^{(j)}. \eqno(12)
$$
The errors in Eqs.~(8) and (10) are again $O(1/n^2)$ and may therefore be neglected for sufficiently large $n$. 

It is clear from these equations that the potential and kinetic energies are obtained from estimators which do not involve any reference to the ring polymer momenta, and indeed one can easily integrate out the momenta from Eqs.~(8) and (10) to leave purely configurational averages. The momenta were originally introduced by Parrinello and Rahman\cite{Parrinello82,Parrinello84} as a sampling tool to facilitate the exploration of configuration space by molecular dynamics. As we have written it, the free ring polymer Hamiltonian $H^0_n({\bf p},{\bf q})$ in Eq.~(5) corresponds to one particular choice of the Parrinello-Rahman mass matrix in which the physical particle mass is assigned to each ring polymer bead. This is the choice that is used in the ring polymer molecular dynamics (RPMD) approximation to real time quantum correlation functions,\cite{Craig04,Braams06} although of course real-time properties are immediately lost as soon as any thermostat is switched on, as we shall do in this paper.

The PIMD method is based on the observation that, since the classical ring polymer trajectory
$$
\dot{\bf p}_t = -{\partial H_n({\bf p}_t,{\bf q}_t)\over\partial {\bf q}_t} \quad
\hbox{and} \quad
\dot{\bf q}_t = +{\partial H_n({\bf p}_t,{\bf q}_t)\over\partial {\bf p}_t} \eqno(13)
$$
conserves both the Boltzmann factor 
$
e^{-\beta_nH_n({\bf p}_t,{\bf q}_t)}=e^{-\beta_nH_n({\bf p}_0,{\bf q}_0)}
$
and the phase space volume element
$
d^f{\bf p}_t d^f{\bf q}_t = d^f{\bf p}_0d^f{\bf q}_0
$,
Eqs.~(8) and~(10) can be re-written as
$$
\left<A\right> \simeq {1\over (2\pi\hbar)^fZ}\int d^f{\bf p}_0\int d^f{\bf q}_0\,
e^{-\beta_nH_n({\bf p}_0,{\bf q}_0)}{\cal A}_n({\bf r}_t). \eqno(14)
$$
This implies that static equilibrium properties such as the potential and kinetic energies can be obtained by time-averaging along ring polymer trajectories whose initial conditions are sampled from the Boltzmann distribution
$$
\rho({\bf p}_0,{\bf q}_0) = {1\over (2\pi\hbar)^fZ}e^{-\beta_nH_n({\bf p}_0,{\bf q}_0)}.
\eqno(15)
$$
Alternatively, and more efficiently, one can combine the sampling with the time evolution by attaching a thermostat to the ring polymer dynamics as we shall describe below. Before we do this, however, let us first explain how to integrate the equations of motion in Eq.~(13) in the absence of a thermostat.

\subsection{Ring polymer time evolution}

A convenient way to integrate Eq.~(13) is based on the splitting of the Hamiltonian in Eq.~(4) into a sum of a free ring polymer part $H_n^0({\bf p},{\bf q})$ and a potential energy part $V_n({\bf q})$. This suggests combining the exact evolutions generated by these two parts of the Hamiltonian in a symplectic integration scheme in which the phase space density evolves under the influence of the symmetric split operator propagator
$$
e^{-\Delta tL} \simeq e^{-(\Delta t/2)L_V}e^{-\Delta t L_0}e^{-(\Delta t/2)L_V},
\eqno(16)
$$
where $L=L_0+L_V$ is the Liouvillian associated with $H_n({\bf p},{\bf q})$ and $L_0$ and $L_V$ are those associated with $H_n^0({\bf p},{\bf q})$ and $V_n({\bf q})$.

The exact evolution generated by $H_n^0({\bf p},{\bf q})$ is simplified by transforming the ring polymer from the bead representation to the normal mode representation,
$$
\tilde{p}_i^{(k)} = \sum_{j=1}^n p_i^{(j)}C_{jk}
\quad\hbox{and}\quad
\tilde{q}_i^{(k)} = \sum_{j=1}^n q_i^{(j)}C_{jk}, \eqno(17)
$$
where in the case of even $n$ the elements of the orthogonal transformation matrix C are
$$
C_{jk} = \begin{cases}
\sqrt{1/n}, & k=0 \cr
\sqrt{2/n}\,\cos\left({2\pi jk/n}\right), & 1\le k\le n/2-1 \cr
\sqrt{1/n}\,(-1)^j, & k=n/2 \cr
\sqrt{2/n}\,\sin\left({2\pi jk/n}\right), & n/2+1\le k\le n-1.
\end{cases} \eqno(18)
$$
In the normal mode representation, $H_n^0({\bf p},{\bf q})$ becomes
$$
H_n^0({\bf p},{\bf q}) = \sum_{i=1}^{N}\sum_{k=0}^{n-1}
\left({[\tilde{p}_i^{(k)}]^2\over 2m_i}+{1\over 2}m_i\omega_k^2[\tilde{q}_i^{(k)}]^2\right),
\eqno(19)
$$
with
$$
\omega_k=2\omega_n\sin\left(k\pi/n\right). \eqno(20)
$$
In view of this, the algorithm implied by Eq.~(16) for integrating the equations of motion in Eq.~(13) through a time interval $\Delta t$ is as follows:
$$
p_{i}^{(j)} \leftarrow p_i^{(j)}-{\Delta t\over 2}{\partial V(q_1^{(j)},\ldots,q_{N}^{(j)})\over \partial q_{i}^{(j)}}, \eqno(21)
$$
$$
\tilde{p}_{i}^{(k)} \leftarrow \sum_{j=1}^n p_{i}^{(j)}C_{jk}, \quad
\tilde{q}_{i}^{(k)} \leftarrow \sum_{j=1}^n q_i^{(j)}C_{jk}, \eqno(22)
$$
$$
\begin{pmatrix}
\tilde{p}_{i}^{(k)} \cr \tilde{q}_{i}^{(k)}
\end{pmatrix} 
\leftarrow
\begin{pmatrix}
\cos \omega_k\Delta t & -m_i\omega_k\sin \omega_k\Delta t \cr
[1/m_i\omega_k]\sin \omega_k\Delta t & \cos \omega_k\Delta t \cr
\end{pmatrix}
\begin{pmatrix}
\tilde{p}_{i}^{(k)} \cr \tilde{q}_{i}^{(k)}
\end{pmatrix}, \eqno(23)
$$
$$
{p}_{i}^{(j)} \leftarrow \sum_{k=0}^{n-1} C_{jk}\,\tilde{p}_i^{(k)}, \quad
{q}_{i}^{(j)} \leftarrow \sum_{k=0}^{n-1} C_{jk}\,\tilde{q}_i^{(k)}, \eqno(24)
$$
$$
p_{i}^{(j)} \leftarrow p_i^{(j)}-{\Delta t\over 2}{\partial V(q_1^{(j)},\ldots,q_{N}^{(j)})\over \partial q_{i}^{(j)}}. \eqno(25)
$$

The first step in this algorithm is an exact evolution of the ring polymer momenta through a time interval $\Delta t/2$ under the influence of the Hamiltonian $V_n({\bf q})$. The second is a transformation from the bead representation to the normal mode representation. The third is an exact evolution of the ring polymer coordinates and momenta through a time interval $\Delta t$ under the influence of the free ring polymer Hamiltonian $H_n^0({\bf p},{\bf q})$. The fourth is a transformation from the normal mode representation back to the bead representation, and the fifth is a further evolution of the ring polymer momenta through a time interval $\Delta t/2$ under the influence of the Hamiltonian $V_n({\bf q})$. 

Because the algorithm consists of a sequence of exact evolutions under the influence of approximate Hamiltonians, it is exactly symplectic,\cite{Sanz-Serna94,Leimkuhler04} which implies that the ring polymer phase space volume will be conserved for any time step $\Delta t$. However, the algorithm will not be very accurate (for example, the ring polymer Hamiltonian will not be very well conserved) unless $\Delta t$ is sufficiently small. When there is only one ring polymer bead, the transformations in Eqs.~(22) and~(24) become redundant, and Eqs.~(21), (23) and (25) reduce to the standard (second order) velocity Verlet method\cite{Verlet67} for integrating a classical trajectory. It is clear from Eq.~(16) that the algorithm remains accurate to $O(\Delta t^2)$ for any number of ring polymer beads. Note also that, in view of the nature of the orthogonal transformation matrix C in Eq.~(18), the transformations in Eqs.~(22) and~(24) can be done using fast Fourier transform routines. These are so efficient that we have not bothered to minimize the number of transformations between the bead and normal mode representations in the implementation of the algorithm we have given above and shall also not bother to do so in the implementation of the thermostats described below.\cite{Note-on-Fourier-transforms}

\subsection{A path integral Langevin equation thermostat}

Bussi and Parrinello have recently explained how a simple (white noise) Langevin thermostat can be combined with the velocity Verlet algorithm to give an efficient sampling of the canonical distribution in classical statistical mechanics.\cite{Ricci03,Bussi07b} Since PIMD is simply classical molecular dynamics in an extended phase space (albeit with a canonical distribution at $n$ times the physical temperature), and since the algorithm we have given in Eqs.~(21) to~(25) is a direct generalization of the velocity Verlet algorithm, it is straightforward to adapt this Langevin thermostat to the present context and use it to thermostat PIMD.

This can be done in two distinct ways, depending on whether one chooses to thermostat the ring polymer beads or the normal modes. Choosing the latter for reasons that will become apparent below, one replaces Eq.~(16) with
$$
e^{-\Delta tL} \simeq e^{-(\Delta t/2)L_{\gamma}}e^{-(\Delta t/2)L_V}e^{-\Delta t L_0}e^{-(\Delta t/2)L_V}e^{-(\Delta t/2)L_{\gamma}},
\eqno(26)
$$
where $L_{\gamma}$ is the part of the Liouvillian $L=L_0+L_V+L_{\gamma}$ in the Fokker-Planck equation for the Langevin phase space density that introduces the friction and the thermal noise.\cite{Bussi07b} Bussi and Parrinello have shown that this is equivalent to adding the following algorithmic steps both before and after Eqs.~(21) to (25):\cite{Bussi07b}
$$
\tilde{p}_{i}^{(k)} \leftarrow \sum_{j=1}^n p_i^{(j)}C_{jk}, \eqno(27)
$$
$$
\tilde{p}_i^{(k)} \leftarrow c_1^{(k)}\tilde{p}_i^{(k)}+\sqrt{m_i\over \beta_n}c_2^{(k)}\xi_i^{(k)}, \eqno(28)
$$
$$
p_i^{(j)} \leftarrow \sum_{k=0}^{n-1} C_{jk}\,\tilde{p}_i^{(k)}. \eqno(29)
$$
Here $\xi_i^{(k)}$ is an independent Gaussian number (a normal deviate with zero mean and unit variance) that is different for each physical degree of freedom, each ring polymer normal mode, and each invocation of Eq.~(28), and the coefficients $c_1^{(k)}$ and $c_2^{(k)}$ are\cite{Bussi07b}
$$
c_1^{(k)} = e^{-(\Delta t/2)\gamma^{(k)}}, \eqno(30)
$$
$$
c_2^{(k)} = \sqrt{1-[c_1^{(k)}]^2}. \eqno(31)
$$

All that remains to complete the specification of this thermostat is to specify the normal mode friction coefficients $\gamma^{(k)}$. The advantage of working in the normal mode representation is that these friction coefficients can be chosen to give an {\em optimal} sampling of the canonical distribution for the free ring polymer, in a sense that we shall now explain. 

In the normal mode representation, the Langevin dynamics of each mode of the free ring polymer is that of an uncoupled harmonic oscillator, 
$$
\eqalign{
{d\over dt}\tilde{q}_{i}^{(k)} &= {\tilde{p}_i^{(k)}\over m_i} \cr
{d\over dt}\tilde{p}_{i}^{(k)} &= -m_i\omega_k^2\tilde{q}_i^{(k)}-\gamma^{(k)}\tilde{p}_i^{(k)}+\sqrt{2m_i\gamma^{(k)}\over\beta_n}\xi_i^{(k)}(t), \cr} \eqno(32)
$$
where $\xi_i^{(k)}(t)$ represents an uncorrelated, Gaussian-distributed random force with unit variance and zero mean [$\bigl<\xi_i^{(k)}\bigr>=0$ and $\bigl<\xi_i^{(k)}(0)\xi_i^{(k)}(t)\bigr>=\delta(t)$.] In view of this, the optimum friction coefficient $\gamma^{(k)}$ will be that which gives the smallest autocorrelation time of the harmonic oscillator Hamiltonian 
and thus the most rapid (Boltzmann-weighted) exploration of free ring polymer energy shells. The relevant autocorrelation time
$$
\tau_H = {1\over \left<H^2\right>-\left<H\right>^2}\int_0^{\infty} \left<(H(0)-\left<H\right>)(H(t)-\left<H\right>)\right>\,dt
\eqno(33)
$$
where
$$
H = {[\tilde{p}_i^{(k)}]^2\over 2m_i}+{1\over 2}m_i\omega_k^2[\tilde{q}_i^{(k)}]^2
\eqno(34)
$$
can be worked out analytically for the Langevin dynamics in Eq.~(32) and is\cite{Zwanzig01,Gardiner03}
$$
\tau_H = {1\over\gamma^{(k)}}+{\gamma^{(k)}\over 4\omega_k^2}. \eqno(35)
$$
The optimum value of $\gamma^{(k)}$ for the excited $(k>0)$ modes of the ring polymer is therefore $\gamma^{(k)} = 2\omega_k$. However, this prescription cannot be used to thermostat the centroid mode, as it gives $\gamma^{(0)}=2\omega_0=0$ (purely Hamiltonian dynamics without a thermostat). We shall therefore define a separate thermostat time constant $\tau_0$ for the centroid mode and specify the normal mode friction coefficients in Eq.~(30) as follows:
$$ 
\gamma^{(k)} =
\begin{cases}
1/\tau_0, & k=0 \cr
2\omega_k, & k>0. \cr
\end{cases}
\eqno(36)
$$

With this specification, the algorithm in Eqs.~(27) to~(29) provides a very simple way to thermostat a PIMD simulation that only requires a single input parameter ($\tau_0$) and that has been tuned to sample the internal modes of the ring polymer as efficiently as possible. The simplicity arises because the tuning is based on the frequencies of the {\em free} ring polymer internal modes, which are known analytically and are independent of the interactions in the system (and therefore transferrable from one system to another). The price to be paid for this is that the optimum friction coefficients for the free ring polymer may not be quite the same as those for the interacting ring polymer, although we would expect them to be very similar for the highest frequency internal modes which are well separated from the vibrations of the physical system. Just how well this \lq\lq path integral Langevin equation" (PILE) thermostat works in practice will be investigated for two different systems in Sec.~III.

\subsection{A generalized Langevin equation thermostat}

Another recent development in thermostatting has been the construction of colored-noise Langevin thermostats for variety of different problems, ranging from separately thermostatting the electrons and the ions in Car-Parrinello molecular dynamics\cite{Ceriotti09a} to efficiently generating configurations consistent with the quantum mechanical canonical ensemble without introducing any path integral beads.\cite{Ceriotti09b} The key point here is that the generalized Langevin equation (GLE) has sufficient flexibility in its memory kernel and its colored noise to be optimized for a variety of different applications.

In the present PIMD context, we would like to choose the colored noise so as to efficiently thermostat a very wide range of frequencies, including both the physical vibrations of the system and the high frequency ring polymer internal modes. That it is possible to construct such a GLE thermostat is shown in Fig.~1, which compares the sampling efficiency of a colored noise thermostat with that of a white noise thermostat with a friction coefficient of $\gamma=\omega_0$. The sampling efficiency is defined\cite{Ceriotti10} as $\kappa(\omega)=[\omega\tau_V(\omega)]^{-1}$, where $\tau_V(\omega)$ is the autocorrelation time for the potential energy of a harmonic oscillator with frequency $\omega$. This provides an indication of how efficiently the thermostat explores the thermally accessible configuration space of a vibration with this frequency. The white noise thermostat is optimally efficient ($\kappa=1$) when $\omega=\omega_0$, but its efficiency decreases linearly for higher and lower values of $\omega$ such that an efficiency of $\kappa>0.2$ is only obtained for a frequency range spanning two orders of magnitude. By contrast,  the colored noise  thermostat achieves $\kappa>0.2$ for a frequency range spanning more than four orders of magnitude. In a 32-bead path integral simulation of liquid water at 300 K, this should be enough to sample everything from the highest frequency internal mode of the ring polymer at $2n/\beta hc\simeq 13,300$ cm$^{-1}$ down to the diffusive modes of the liquid at frequencies as low as 1 cm$^{-1}$.

The colored noise thermostat in Fig.~1 is one of a family of thermostats that have recently been constructed\cite{Ceriotti10} by exploiting the equivalence\cite{Zwanzig01} between the non-Markovian dynamics of the GLE and Markovian dynamics in a higher dimensional space. These thermostats have a simple matrix form that is straightforward to implement on top of the ring polymer evolution algorithm in Eqs.~(21) to~(25). Choosing to work in the bead representation rather than the normal mode representation, since there is no advantage in this case in making the transformation to normal modes, one simply replaces Eqs.~(27) to~(29) with\cite{Ceriotti10}
$$
{\bf p}_i^{(j)} \leftarrow {\bf C}_1{\bf p}_i^{(j)}+\sqrt{m_i\over\beta_n}{\bf C}_2\,\bm{\xi}_i^{(j)}. \eqno(37)
$$
Here $\bm{\xi}_{i}^{(j)}$ is a vector of $n_s+1$ independent Gaussian numbers, 
$$
{\bf p}_i^{(j)} = \begin{pmatrix}
p_i^{(j)} \cr {\bf s}_i^{(j)}\cr \end{pmatrix} \eqno(38)
$$
is a vector containing the momentum of bead $j$ and $n_s$ auxilliary momenta ${\bf s}_i^{(j)}$, and 
$$
{\bf C}_1 = e^{-(\Delta t/2)\bm{\gamma}^T}, \eqno(39)
$$
and
$$
{\bf C}_2^T{\bf C}_2 = {\bf I}-{\bf C}_1^T{\bf C}_1, \eqno(40)
$$
are the appropriate matrix generalisations of Eqs.~(30) and~(31).

Given a suitable friction matrix $\bm{\gamma}$, the implementation of this algorithm is as follows. One begins by constructing ${\bf C}_1$ from Eq.~(39) and ${\bf C}_2$ from Eq.~(40). ${\bf C}_1$ can be constructed by combining a low-order Taylor series expansion of $e^{-(\Delta t/2^{P+1})\bm{\gamma}^T}$ with $P$ matrix-squaring operations, and ${\bf C}_2$ can be constructed by Cholesky decomposition of ${\bf I}-{\bf C}_1^T{\bf C}_1$. These matrices are then stored for use in each iteration of Eq.~(37), which requires two matrix-vector multiplications and the generation of $n_s+1$ normal deviates for each degree of freedom $i$ and each ring polymer bead $j$. 

In most large-scale applications, the effort required by this algorithm will be negligible compared with the evaluation of the forces in Eqs.~(21) and~(25). If it is not, one can exploit the fact that the canonical distribution is invariant under the action of Eq.~(37) for any time step $\Delta t$, and only apply the thermostat once in every $m$ time steps with $\Delta t$ in Eq.~(39) replaced by $m\Delta t$.\cite{Ceriotti10} With this modification, the algorithm clearly provides a computationally very tractable way to thermostat a PIMD simulation.

All that remains is to specify the friction matrix $\bm{\gamma}$ in Eq.~(39). In the calculations described below, we used $n_s=4$ and
$$
\bm{\gamma}=\omega_0\,{\bf A}, \eqno(41)
$$
where the elements of the $(5\times 5)$ matrix ${\bf A}$ are given in Table I. These matrix elements were obtained in an automated procedure\cite{Ceriotti10} designed to optimize the efficiency $\kappa(\omega)$ of the thermostat, which we have already compared with that of the corresponding white noise Langevin thermostat ($n_s=0$, $\gamma=\omega_0$) in Fig.~1. In that comparison, $\omega_0$ was chosen arbitrarily, but in actual applications it is a physical input parameter that can be used to tune the response of the thermostat to a particular frequency range, which one sees to be roughly $0.01\,\omega_0\le \omega \le 100\,\omega_0$ from the results in Fig.~1. One can of course equally well replace this input parameter with a thermostat time constant $\tau_0=1/2\omega_0$, which we shall do in our comparison of various thermostats in Sec.~III.

\begin{table}
\begin{center}
\caption{Dimensionless elements $A_{ij}$ of the GLE friction matrix ${\bf A}$ in Eq.~(41). This matrix was constructed using the options $n_s=4$, range = 10$^4$, and \lq\lq fit for potential energy", using the freely available software on the website http://gle4md.berlios.de/.}
\bigskip
\begin{tabular}{ccc} \hline
\phantom{xx} $i$\phantom{xx} &\phantom{xx} $j$\phantom{xx} & $A_{ij}$ \\
\hline\\
1 & 1 & 2.468046483820e+1 \\
1 & 2 & 3.618484148135e-2  \\
1 & 3 & 1.529754837748e+0 \\   
1 & 4 & -4.832976901522e+0 \\
1 & 5 &  3.075592122514e+1 \\ 
2 & 1 & -3.690906142217e-2 \\
2 & 2 &  1.140757569304e-5 \\
2 & 3 &  9.580998002948e-2 \\
2 & 4 &  -2.633785831010e-2 \\
2 & 5 &  5.628596350432e-2 \\
3 & 1 &  -1.967695128248e+0 \\
3 & 2 &  -9.580998002948e-2 \\
3 & 3 &  1.803797247061e-1 \\
3 & 4 &  6.834981703810e-1 \\
3 & 5 &  -1.326536043516e+0 \\
4 & 1 &  -1.376606646573e+0 \\
4 & 2 &  2.633785831010e-2 \\
4 & 3 &   -6.834981703810e-1 \\
4 & 4 &   3.538593762043e+0 \\
4 & 5 &  1.527314768745e+0 \\
5 & 1 &  2.893495089306e+1 \\
5 & 2 &   -5.628596350432e-2 \\
5 & 3 &  1.326536043516e+0 \\
5 & 4 &  -1.527314768745e+0 \\
5 & 5 &  4.108827095695e+1 \\ \\  
\hline
\end{tabular}
\end{center}
\end{table}

\begin{figure*}
\includegraphics{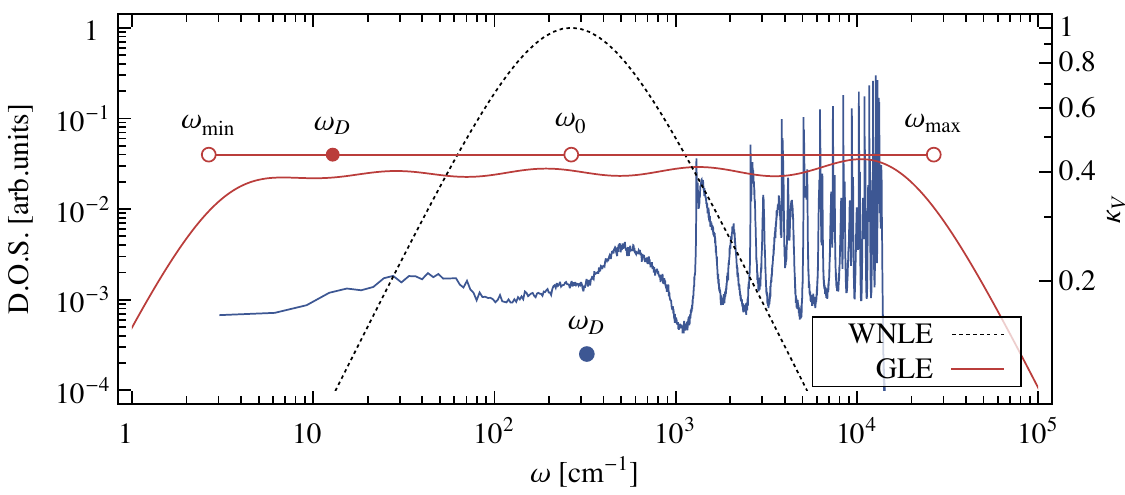}
\caption{The vibrational spectrum from a path integral molecular dynamics simulation of a flexible water model (blue). In order to include the internal modes of the ring polymer, the density of states has been computed from the velocity autocorrelation function of the individual ring polymer beads. This vibrational spectrum is compared with the harmonic-limit sampling efficiency $\kappa_{V}$ of a white-noise Langevin thermostat (dashed) and an optimal GLE thermostat (red). The latter has been optimized so as to give constant sampling efficiency over a frequency range spanning four orders of magnitude, centered geometrically on $\omega_{0}=1/2\tau_{0}$. The diffusion coefficient for the water model, and that computed for the GLE dynamics in the free-particle limit, are also shown in terms of the ``diffusion frequency'' $\omega_{D}=k_{\rm B}T/mD$.}
\end{figure*}

For the sake of comparison, we will also perform some calculations in Sec.~III with a white noise Langevin equation thermostat with friction $\gamma=\omega_0=1/2\tau_0$. This white noise Langevin scheme is closely related to an Andersen thermostat in which the momenta are resampled with a probability $\gamma \Delta t$ after each time step $\Delta t$. The two methods have a similar sampling efficiency in all the examples we have studied, and for this reason the results with the Andersen thermostat will not be reported.

\subsection{The Nos\'e-Hoover chain thermostat}

As discussed in the Introduction, the gold standard against which to compare the above stochastic PILE and GLE thermostats is a deterministic Nos\'e-Hoover chain\cite{Martyna92} applied separately to each physical degree of freedom and each ring polymer bead (or ring polymer normal mode).\cite{Tuckerman93}

One convenient way to do this that reduces to a recommended\cite{Martyna96} operator splitting in the classical limit is to replace Eq.~(16) with
$$
e^{-\Delta t L} \simeq e^{-(\Delta t/2)L_{NHC}}e^{-(\Delta t/2)L_V}e^{-\Delta t L_0}e^{-(\Delta t/2)L_V}e^{-(\Delta t/2)L_{NHC}}, \eqno(42)
$$
where $L_{NHC}$ is the part of the Liouvillian $L=L_0+L_V+L_{NHC}$ for the extended system dynamics that involves the Nos\'e-Hoover chain variables.\cite{Martyna96} Choosing to work in the normal mode representation as in the case of the PILE thermostat, the net effect of this operator splitting is that Eq.~(28) is replaced by the evolution of the following system of non-linear differential equations\cite{Martyna92} through a time interval of $\Delta t/2$:
$$
{d\over dt}\tilde{p}_i^{(k)} = -\tilde{p}_i^{(k)}{\pi_{i,1}^{(k)}\over Q^{(k)}}, \eqno(43)
$$
$$
{d\over dt}\pi_{i,1}^{(k)} = \left({[\tilde{p}_i^{(k)}]^2\over m_i}-{1\over\beta_n}\right)-\pi_{i,1}^{(k)}{\pi_{i,2}^{(k)}\over Q^{(k)}}, \eqno(44)
$$
$$
{d\over dt}\pi_{i,l}^{(k)} = \left({[\pi_{i,l-1}^{(k)}]^2\over Q^{(k)}}-{1\over\beta_n}\right)-\pi_{i,l}^{(k)}{\pi_{i,l+1}^{(k)}\over Q^{(k)}}, \eqno(45)
$$
$$
{d\over dt}\pi_{i,L}^{(k)} = \left({[\pi_{i,L-1}^{(k)}]^2\over Q^{(k)}}-{1\over\beta_n}\right), \eqno(46)
$$
$$
{d\over dt}\eta_{i,l}^{(k)} = {\pi_{i,l}^{(k)}\over Q^{(k)}}. \eqno(47) 
$$
Here $\pi_{i,l}^{(k)}$ and $\eta_{i,l}^{(k)}$ are the momentum and position variables of the Nos\'e-Hoover chain attached to the $k$-th normal mode of the ring polymer in the $i$-th degree of freedom, for $l=1,\ldots,L$. 

The system of differential equations in Eqs.~(43) to (47) has been discussed extensively in the literature on Nos\'e-Hoover chains and numerous numerical methods have been proposed for integrating it.\cite{Tuckerman93,Martyna96,Jang97} The accuracy of the integration through each time interval $\Delta t/2$ can be checked separately for each $i$ and $k$ by monitoring the locally conserved quantity
$$
{H'}_i^{(k)} = {[\tilde{p}_i^{(k)}]^2\over 2m_i}+\sum_{l=1}^L \left({[\pi_{i,l}^{(k)}]^2\over 2Q^{(k)}}+{\eta_{i,l}^{(k)}\over\beta_n}\right), \eqno(48)
$$ 
and the accuracy of the overall symmetric split operator propagator in Eq.~(42) can be checked by monitoring the globally conserved quantity
$$
H'_n = H_n({\bf p},{\bf q})+\sum_{i=1}^{N}\sum_{k=0}^{n-1}\sum_{l=1}^L \left({[\pi_{i,l}^{(k)}]^2\over 2Q^{(k)}}+{\eta_{i,l}^{(k)}\over\beta_n}\right). \eqno(49)
$$

In order to compare this thermostat with the stochastic thermostats introduced above, we need to specify the normal mode thermostat masses $Q^{(k)}$. These play an analogous role to the normal mode friction coefficients $\gamma^{(k)}$ in the PILE thermostat introduced in Sec.~II.C.  The Nos\'e-Hoover chain masses that give the most efficient sampling of the canonical distribution for the free ring polymer are\cite{Martyna92} $Q^{(k)}=1/\beta_n\omega_k^2$. However, we cannot use this prescription to thermostat the centroid mode, as $\omega_0=0$. As in the case of the PILE thermostat, we can get around this difficulty by defining a separate thermostat time constant $\tau_0$ for the centroid mode and replacing $\omega_0$ with $1/2\tau_0$, so that the normal mode masses that appear in Eqs.~(43) to~(47) are specified as follows:
$$
Q^{(k)} = \begin{cases}
4\tau_0^2/\beta_n, & k=0 \cr
1/\beta_n\omega_k^2, & k>0. \cr
\end{cases} \eqno(50)
$$

This completes the specification of the NHC thermostat that we shall use to compare with the stochastic PILE and GLE thermostats in Sec.~III. Several other groups have proposed slightly different implementations of the NHC thermostat for PIMD simulations, but we would not expect the differences to have any major impact on our results. For example, the original implementation by Tuckerman {\em et al.}\cite{Tuckerman93} involved transforming to \lq\lq staging" variables rather than normal mode variables, and employed a different operator splitting than the one we have used in Eq.~(42). Staging variables are a valid alternative to normal mode variables and there are a number of different ways in which one can do the operator splitting, all of which should in principle converge on the same result. We have used normal mode variables here to make the connection with the free ring polymer evolution algorithm in Eqs.~(21) to~(25),  and chosen the operator splitting in Eq.~(42) to emphasize the connection with the PILE thermostat in Eqs.~(27) to~(29). A general feature of NHC thermostats for path integrals is that they require the solution of a large number of systems of non-linear differential equations of the form in Eqs.~(43) to~(47). Since these differential equations must be solved numerically these thermostats are more complicated than the stochastic thermostats we have described above.

\subsection{\lq\lq Global" versus \lq\lq local" thermostatting}

All of the thermostats we have discussed so far are \lq\lq local", in the sense that each degree of freedom and each ring polymer bead (or normal mode) is separately thermostatted. This is the simplest way to ensure that the kinetic and potential energies of every single particle in the system are thermalized as rapidly as possible, and it is therefore likely to be the most efficient way to sample local properties such as the energy of an interstitial hydrogen atom in palladium. On the other hand, there are many properties that are sensitive to the slow, collective motions of all of the atoms in the system, such as the dielectric constant of liquid water. Local thermostats will almost certainly be too aggressive to give an efficient sampling of these collective properties because with even a moderate amount of friction they tend to to inhibit the Hamiltonian diffusion. There has therefore been a renewed interest recently in the development of \lq\lq global" thermostats that are attached to the system in a more gentle way and have a less disruptive effect on the diffusion.

In particular, Bussi {\em et al.}\cite{Bussi07a,Bussi08} have derived a global version of the finite time step Langevin propagator in Eq.~(28) that acts as a thermostat on the kinetic energy of all $N$ degrees of freedom rather than on each degree of freedom separately. When implementing this global thermostat in a path integral context, it is desirable to apply it only to the centroid mode, and to leave more aggressive local thermostats attached to the less ergodic\cite{Hall84} excited ring polymer internal modes. In practice, this amounts to replacing just the centroid $(k=0)$ component of Eq.~(28) with the following stochastic velocity rescaling algorithm:\cite{Bussi07a,Bussi08}
$$
\tilde{p}_i^{(0)}\leftarrow \alpha\,\tilde{p}_i^{(0)}, \eqno(51)
$$
where
$$
\alpha^2 = c+{(1-c)\bigl(\bigl[\xi_1^{(0)}\bigr]^2+\sum_{i=2}^{N} \bigl[\xi_i^{(0)}\bigr]^2\bigr)\over 2\beta_nK}
+2\xi_1^{(0)}\sqrt{c(1-c)\over 2\beta_nK}, \eqno(52)
$$
and
$$
{\rm sign}\left[\alpha\right] = {\rm sign}\left[\xi_1^{(0)}+\sqrt{2\beta_nKc\over (1-c)}\right],
\eqno(53)
$$
with
$$
K = \sum_{i=1}^{N} {[\tilde{p}_i^{(0)}]^2\over 2m_i}, \eqno(54)
$$
and $c=e^{-\Delta t\gamma^{(0)}}$. We shall call this global version of the PILE thermostat PILE-G and the local version in which the centroid is thermostatted using Eq.~(28) PILE-L in the numerical comparisons presented below.

The Nos\'e-Hoover chain thermostat can also be applied globally and indeed this is how classical canonical ensemble simulations are often carried out.  Choosing again to keep the thermostatting of the excited ring polymer modes local, the modification amounts in this case to replacing the centroid $(k=0)$ components of Eqs.~(43) to~(47) with\cite{Martyna92}
$$
{d\over dt}\tilde{p}_i^{(0)} = -\tilde{p}_i^{(0)}{\pi_{1}^{(0)}\over Q^{(0)}}, \eqno(55)
$$
$$
{d\over dt}\pi_{1}^{(0)} = \left(\sum_{i=1}^{N}{[\tilde{p}_i^{(0)}]^2\over m_i}-{N\over\beta_n}\right)-\pi_{1}^{(0)}{\pi_{2}^{(0)}\over Q^{(0)}}, \eqno(56)
$$
$$
{d\over dt}\pi_{l}^{(0)} = \left({[\pi_{l-1}^{(0)}]^2\over Q^{(0)}}-{1\over\beta_n}\right)-\pi_{l}^{(0)}{\pi_{l+1}^{(0)}\over Q^{(0)}}, \eqno(57)
$$
$$
{d\over dt}\pi_{L}^{(0)} = \left({[\pi_{L-1}^{(0)}]^2\over Q^{(0)}}-{1\over\beta_n}\right), \eqno(58)
$$
$$
{d\over dt}\eta_{l}^{(0)} = {\pi_{l}^{(0)}\over Q^{(0)}}, \eqno(59) 
$$
and to making the appropriate modifications to Eqs.~(48) and~(49); for example Eq.~(48) is replaced by\cite{Martyna92}
$$
{H'}^{(0)} = \sum_{i=1}^{N}{[\tilde{p}_i^{(0)}]^2\over 2m_i}+\sum_{l=1}^L {[\pi_{l}^{(0)}]^2\over 2Q^{(0)}}+{N\eta_{1}^{(0)}\over\beta_n}+\sum_{l=2}^L
{\eta_{l}^{(0)}\over\beta_n}. \eqno(60)
$$ 
Notice in particular that the $N$ separate Nos\'e-Hoover chains in the local thermostat have been replaced with a single chain that is coupled to the total kinetic energy of the ring polymer centroid via Eq.~(56). We shall call this global version of the NHC thermostat NHC-G and the local version in which the centroid is thermostatted using Eqs.~(43) to~(47) 
NHC-L in the comparisons presented below.

One could clearly also imagine trying to develop a global version of the GLE thermostat introduced in Sec.~II.D. However, little benefit would be expected from this generalization, because the capability of the GLE to adapt automatically to the vibrational modes present in the system relies on the fact that an independent thermostat is applied to each degree of freedom. In any event, the two global thermostats we have just described (PILE-G and NHC-G) are sufficient for us to illustrate when it is preferable to use a global thermostat rather than a local thermostat for path integral simulations, as we shall do in the following section.

\section{Results and Discussion}

It is well known that molecular dynamics can be used to efficiently explore regions of phase space where the free energy surface contains local minima separated by barriers comparable to the average thermal energy. Thermostats can further enhance this sampling, particularly in the case of PIMD where poorly ergodic, harmonic necklace modes are present. In some cases inefficient sampling can also arise when a statistically significant portion of phase space consists of regions separated by high free energy barriers. In these situations, which are not the concern of the present work, the fact that PIMD is simply classical molecular dynamics in an extended phase space should in principle enable one to adopt many of the accelerated dynamics techniques that have been developed to speed up the rates of transitions across barriers in classical simulations.

When this issue is not present a quantitative way assess the sampling efficiency is to consider the correlation time of an observable $A$,
$$
\tau_{A} = {1\over \left<{A}^2\right>-\left<{A}\right>^2}\int_0^{\infty} \left<({A}(0)-\left<{A}\right>)({A}(t)-\left<{A}\right>)\right>\,dt, \eqno(61)
$$
in which the angular brackets denote an average of the appropriate path integral estimator ${\cal A}_n({\bf q})$ along a thermostatted PIMD trajectory. This is related to the statistical uncertainty $\epsilon_{A}$ in the expectation value of the observable $\left<{A}\right>$ by
$$
\epsilon_{A} \propto \sqrt{ \frac{\tau_{A}}{t_{\rm sim}}}, \eqno(62)
$$
where $t_{\rm sim}$ is the total simulation time. Therefore, one would like to make $\tau_{A}$ as small as possible so as to reduce the uncertainty in $\left<A\right>$ for a simulation of a given length. In the previous section we have used the correlation time of the total energy for a harmonic oscillator as a tentative measure of the efficiency of sampling for a Langevin equation [see Eqs.~(33) and~(35)]. Here we will instead perform PIMD simulations of two typical condensed-phase systems and compute the correlation times of a number of different physical observables so as to compare the sampling efficiencies of various thermostats in more realistic (anharmonic) situations.

\subsection{Liquid water}

Liquid water is ubiquitous in biological systems and important in many applications in chemistry and materials science. Nuclear quantum effects have a significant influence on its properties and it has therefore been a common target of PIMD studies ever since the first path integral simulations of the liquid were performed in the mid 1980s.\cite{Kuharski85,Wallqvist85} It is also a highly structured liquid in which changes in the local structure depend on complex collective rearrangements of the hydrogen bonding network, which are difficult to sample. Because of this, liquid water is a highly relevant test case for our comparison.

Simulations were performed using the q-TIP4P/F flexible, four-site water potential which has recently been parameterized to reproduce many of the static and dynamical properties of water in path integral simulations.\cite{Habershon09} For each thermostat and value of the time constant $\tau_0$, we ran a trajectory for $12$~ns at a temperature of 298~K and a density of 0.997 g cm$^{-3}$. The simulation box contained $216$ molecules, with each atom described by a $32$ bead ring polymer. The equations of motion were integrated as described in Sec.~II, and the ring-polymer contraction scheme of Markland {\em et al.}\cite{Markland08b,Markland08c} was used to accelerate the evaluation of the long-range electrostatic interactions.

In order to assess the efficiencies of the different thermostats, we first computed the correlation time $\tau_T$ of the centroid virial estimator of the kinetic energy. While for a system of classical particles the kinetic energy is simply distributed according to the Boltzmann law, for quantum systems there is also a position-dependent contribution which requires one to sample the forces experienced by the internal modes of the ring polymer. This contribution is included in the centroid virial estimator in Eq.~(11), which can be written equivalently in the normal mode representation as
$$
{\cal T}_n({\bf q}) ={N\over 2\beta}-{1\over 2n}\sum_{i=1}^N\sum_{k=1}^{n-1} \tilde{q}_{i}^{(k)}\cdot \tilde{F}_i^{(k)}(\bf q), \eqno(63)
$$
where $\tilde{F}_i^{(k)}({\bf q}) = -{\partial V_n({\bf q})/\partial\tilde{q}_i^{(k)}}$ is the force on the $k$-th excited $(k>0)$ normal mode of the ring polymer that arises from the physical interaction potential.

Fig.~2 compares the kinetic energy correlation times $\tau_T$ for a white noise Langevin equation (WNLE) themostat, the GLE themostat, and global and local versions of the PILE and NHC thermostats as a function of the thermostat time constant $\tau_0$. The performance of the simple WNLE thermostat, which yields efficient thermalization in the high friction limit but is very inefficient for large $\tau_0$, shows that effective sampling of ${\cal T}_n({\bf q})$ requires strong coupling to the high frequency necklace modes. This is achieved automatically in the PILE and NHC thermostats, which specifically target the ring polymer normal modes, overcome the ergodicity problems observed in microcanonical PIMD and consistently yield a low value of $\tau_T$. Note that this is true regardless of the method (global or local) that is used to thermostat the centroid, which does not contribute to the kinetic energy estimator in Eq.~(63). The quality of the GLE thermostat also deteriorates more slowly than that of the WNLE thermostat with increasing $\tau_0$, and is constant as long as all of the ring polymer necklace modes are included in the optimal range of fitted frequencies (see Fig.~1).

\begin{figure}
\includegraphics{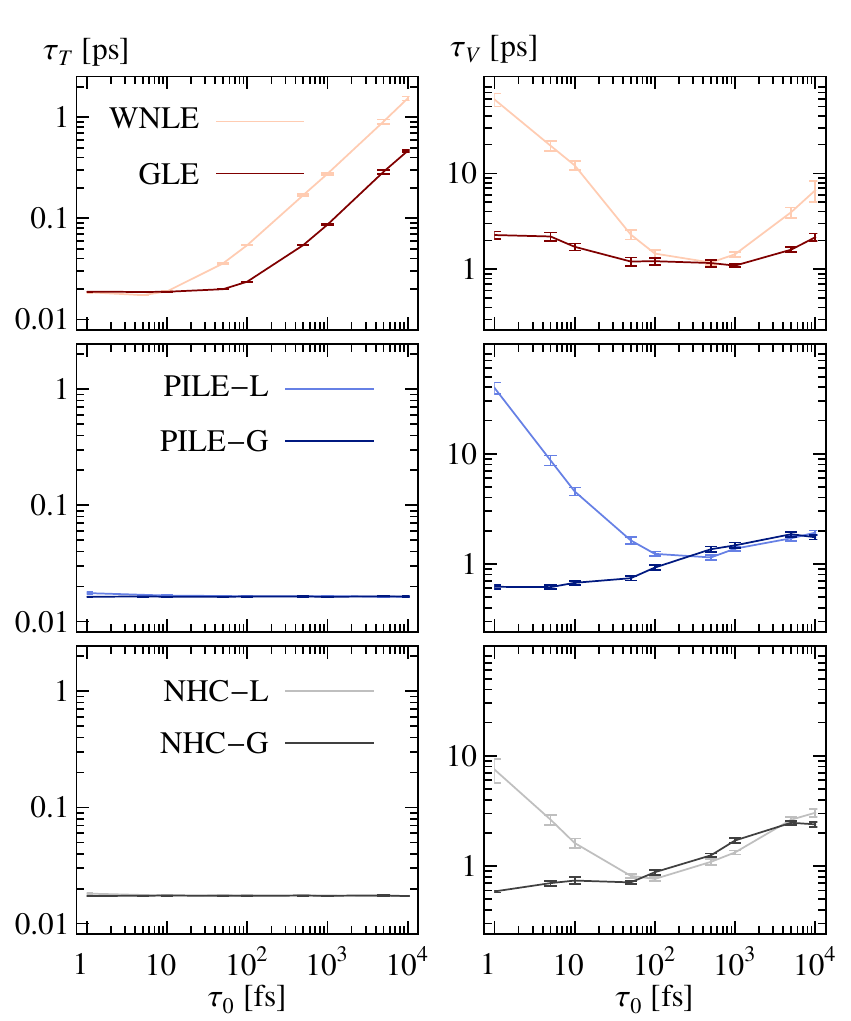}
\caption{
Correlation times for the centroid virial kinetic energy (left) and the potential energy (right) obtained from path integral molecular dynamics simulations of liquid water. The six panels show the dependence of these quantities on the thermostat relaxation time $\tau_0$ for the WNLE and GLE thermostats and for local and global versions of the PILE and NHC thermostats.}
\end{figure}

\begin{figure}
\includegraphics{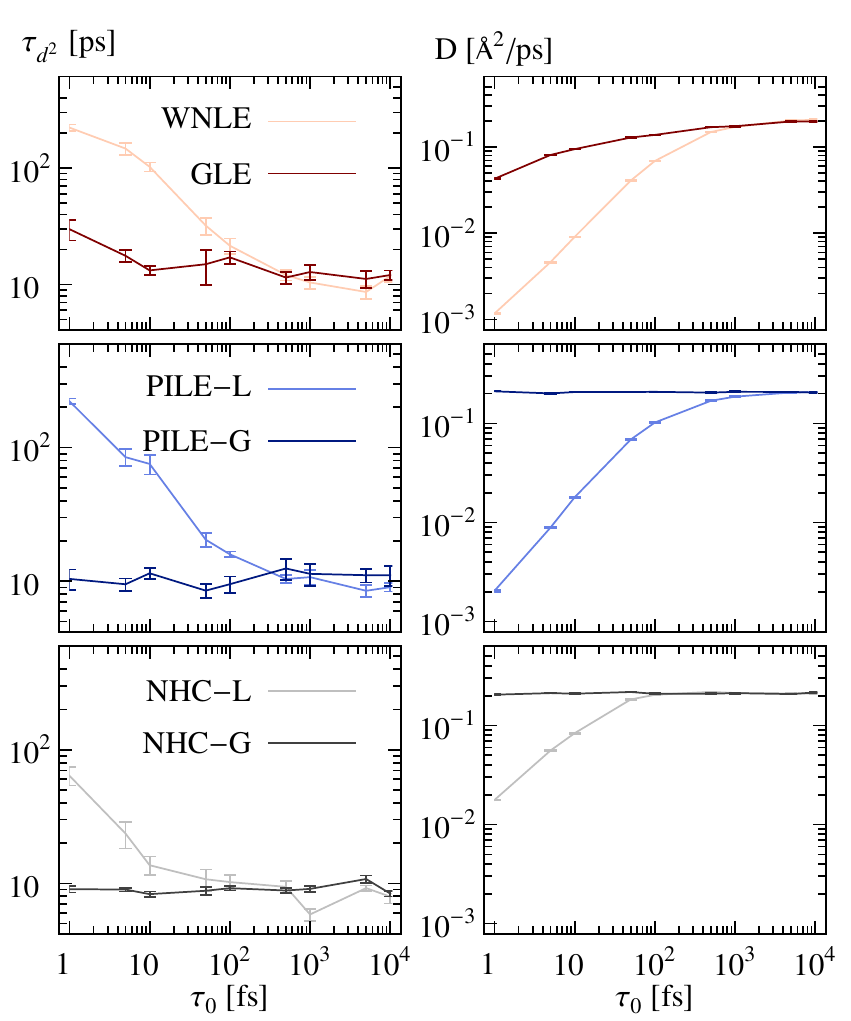}
\caption{
Correlation time for the squared dipole moment of the simulation box (left) and the computed molecular centre-of-mass diffusion coefficient (right) obtained from path integral simulations of liquid water. The six panels show the dependence of these quantities on the thermostat relaxation time $\tau_0$ for the WNLE and GLE thermostats and for local and global versions of the PILE and NHC thermostats.}
\end{figure}

For comparison with $\tau_T$, we also report in Fig.~2 the correlation time of the potential energy, $\tau_V$. This is much more sensitive to the softer modes, and to the diffusive motion of the centroid in particular. Efficiently sampling the potential energy is considerably (at least an order of magnitude) more difficult than sampling the kinetic energy, as can be seen from the correlation times in the figure. This is because one must carefully balance the strong coupling needed to ensure ergodicity of the polymers with the need to avoid overdamping which causes the thermostat to become an additional bottleneck to diffusion. The diffusion bottleneck is quite apparent in Fig.~2 from the significant increase in $\tau_V$ which is observed for all of the local thermostats when $\tau_0$ is decreased below around 100~fs.

A reasonable efficiency in the sampling of the potential energy can be achieved by carefully tuning the simple WNLE thermostat, with the optimal coupling being $\tau_0\simeq 500$~fs. However, tuning the thermostat in this way for every system one might wish to simulate is clearly undesirable and large increases in the potential energy autocorrelation time result from a less than optimal choice of the WNLE time constant. In contrast, the GLE, which also requires no {\it a priori} knowledge of the internal frequencies of the ring polymer, obtains a nearly constant sampling efficiency ($\tau_V\simeq 2$ ps) over the five orders of magnitude of $\tau_0$ we have considered. 

Turning now to the targeted schemes which exploit a knowledge of the ring polymer normal mode frequencies, one sees from Fig.~2 that the potential energy is sampled most efficiently by PILE-G and NHC-G, which use a global thermostat for the centroid. These schemes particularly excel in the case of strong coupling (i.e., for a small relaxation time $\tau_0$), implying that a rapid rescaling of the total (centroid) kinetic energy helps to speed up the canonical sampling of the potential energy. At the same time, since the global thermostats are coupled to the total kinetic energy rather than to individual momenta, the trajectory and the dynamical properties are only slightly disturbed,\cite{Bussi07a} and hence the ability of Hamiltonian dynamics to generate complex, collective rearrangements can be fully exploited. This is clear from the superiority of the global thermostats over the local thermostats for this problem. An important final point from Fig.~2 is that the simple stochastic PILE-G thermostat provides equally rapid sampling of both the kinetic energy and the potential energy of liquid water as the NHC-G thermostat, without the need to evolve any extended variables.

The superiority of global thermostats over local thermostats for liquid water is further supported by the results in Fig.~3, where we show the correlation time of the squared dipole moment of the supercell and the center-of-mass diffusion coefficient obtained from the thermostatted trajectories. The squared dipole moment is related to the dielectric constant and is known to converge slowly as it requires the reorientation of many individual water molecules, which in turn requires a concerted rearrangement of the hydrogen-bonding network. In this case, microcanonical dynamics is extremely effective, and the best sampling is obtained with very weak coupling to the thermostats. For the local WNLE, PILE-L and NHC-L thermostats, there is a clear correspondence in Fig.~3 between the reduction of the diffusion coefficient due to over-damped dynamics and the degradation of $\tau_{d^2}$ as the thermostat time constant $\tau_0$ is decreased. The global PILE-G and NHC-G thermostats, and to a lesser extent also the GLE thermostat, give much more effective sampling of the squared dipole moment for small values of $\tau_0$ and also have a less disruptive effect on the diffusion. The properties of the correlated noise we have used in the GLE are clearly such that its effect on diffusion is much less severe than that of the WNLE,\cite{Ceriotti10} despite its strong coupling to the high frequency necklace modes. As a consequence, its effects are intermediate between those of local and global schemes.

Combining the results in Figs.~2 and~3, we see that the PILE-G and NHC-G thermostats are the most efficient at sampling the properties of liquid water that we have considered, with the GLE also doing rather well (and certainly much better than the simple WNLE). However, one cannot conclude from this that it will always be better to use a global rather than a local thermostatting scheme. Local schemes enforce a canonical distribution on each of the individual components of the momentum, and therefore ensure a homogeneous distribution of energy throughout the system. Global schemes only monitor the overall temperature of the system, and rely on anharmonic coupling between different regions and different frequency ranges to reach equipartition. For these reasons, one should be particularly careful when using global thermostats, as the inefficient sampling of particular internal degrees of freedom may be masked by the very efficient equilibration of the total temperature. A few local properties should therefore also be monitored to ensure that equilibrium has been reached, and in general a local thermostat is to be preferred whenever inhomogeneous or quasi-harmonic problems are treated, as we will demonstrate in the next subsection.

\subsection{Hydrogen in palladium}

The second example system we shall consider is both inhomogeneous and quasi-harmonic: the motion of an atomic hydrogen (H) interstitial in the lattice of palladium (Pd). In this more harmonic system, strong anti-correlations are present in the underdamped limit. In some cases this can be beneficial, and methods have been devised which explicitly enhance anti-correlations.\cite{Adler81} However, our aim here is to estimate how efficiently uncorrelated configurations
are generated by the thermostatted dynamics. An analysis based on correlation times of the form in Eq.~(61) would therefore be misleading, as anti-correlations can significantly reduce $\tau_A$ and mask the presence of very long relaxation times. For this reason, the correlation times reported in this subsection are computed as the integral of the {\em absolute value} of the normalized autocorrelation function,
$$
\tilde{\tau}_{A} = {1\over \left<{A}^2\right>-\left<{A}\right>^2}\int_0^{\infty} \left|\left<({A}(0)-\left<{A}\right>)({A}(t)-\left<{A}\right>)\right>\right|\,dt. \eqno(64)
$$

The calculations used a supercell containing 256 Pd atoms and a single H atom. A 10 bead ring polymer was used for H while, in view of their large mass, the Pd atoms were treated as classical particles. The Pd lattice parameter was 3.89 \AA\ and the temperature was chosen to be $350$~K, where diffusion of the H atom is still relatively fast ($D=0.16$ \AA $^{2}$ ps$^{-1}$). Interactions were modeled by an embedded atom potential.\cite{Zhou08} We performed $8$~ns of simulation with a time step of 0.5~fs for each choice of thermostat and $\tau_0$.

The thermostatting of the classical metal lattice and the H ring polymer were performed separately. The same coupling parameters and thermostatting method were used for the metal and for the centroid of the H atom necklace. With this setup global thermostats on the H atom centroid act on only three degrees of freedom. Nevertheless, significant differences are still observed with respect to fully local schemes.

In Fig.~4 we plot the correlation times of the radius of gyration of the ring polymer and of the H atom kinetic energy. The path integral estimators for both of these quantities depend solely on the internal necklace modes, and the methods which target these modes specifically (PILE and NHC) offer clear advantages. In particular, the radius of gyration of the H atom ring polymer, which is almost completely decoupled from the slow motion of the ring polymer centroid, is seen to be thermostatted equally well by PILE and NHC. This clearly demonstrates that, by targeting the normal modes with either thermostat (stochastic or deterministic), the ergodicity problem of PIMD is completely resolved. The GLE is also seen to give rapid sampling of the H atom radius of gyration and kinetic energy without exploiting any knowledge of the normal mode vibrational frequencies, unless its fitted range is shifted so that it does not encompass all of the relevant spectrum ($\tau_0>100$~fs).

\begin{figure}
\includegraphics{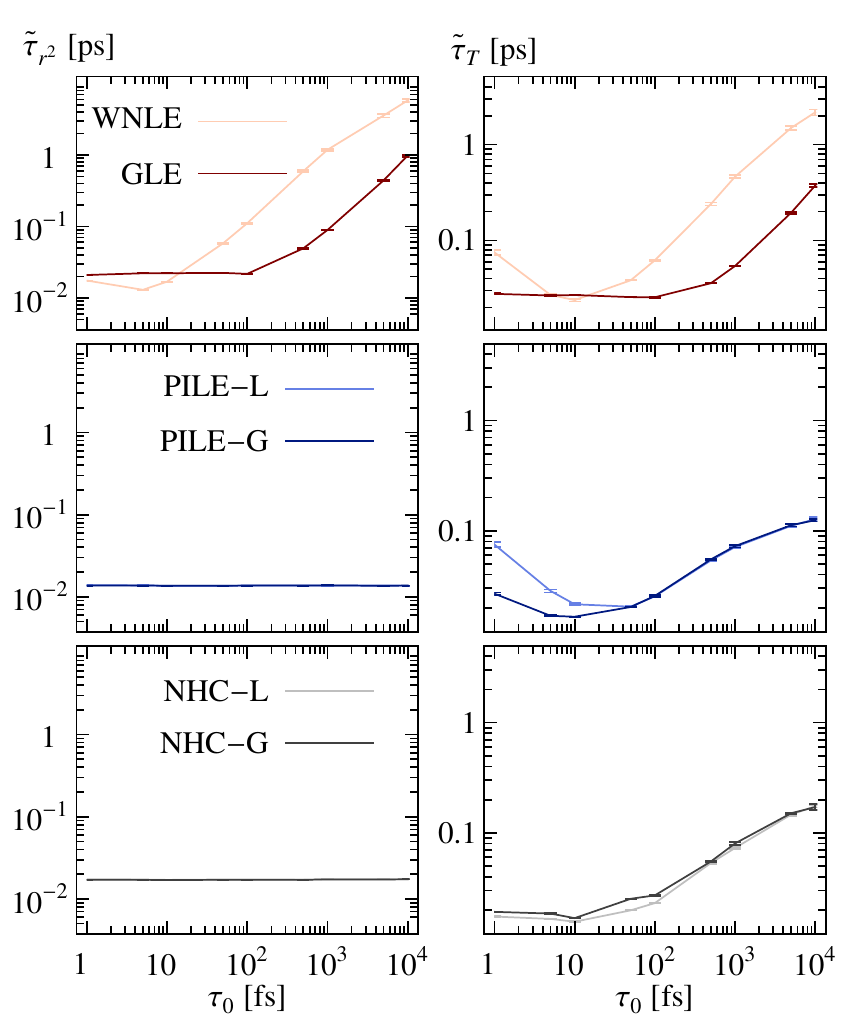}
\caption{
Integrals of the absolute values of the normalized autocorrelation functions $\tilde{\tau}$ (Eq.~64) for the ring polymer radius of gyration (left) and the kinetic energy (right) of a hydrogen atom interstitial in a palladium lattice, obtained from path integral molecular dynamics simulations. The six panels show the dependence of these quantities on the thermostat relaxation time $\tau_0$ for the WNLE and GLE thermostats and for local and global versions of the PILE and NHC thermostats.}
\end{figure}

\begin{figure}
\includegraphics{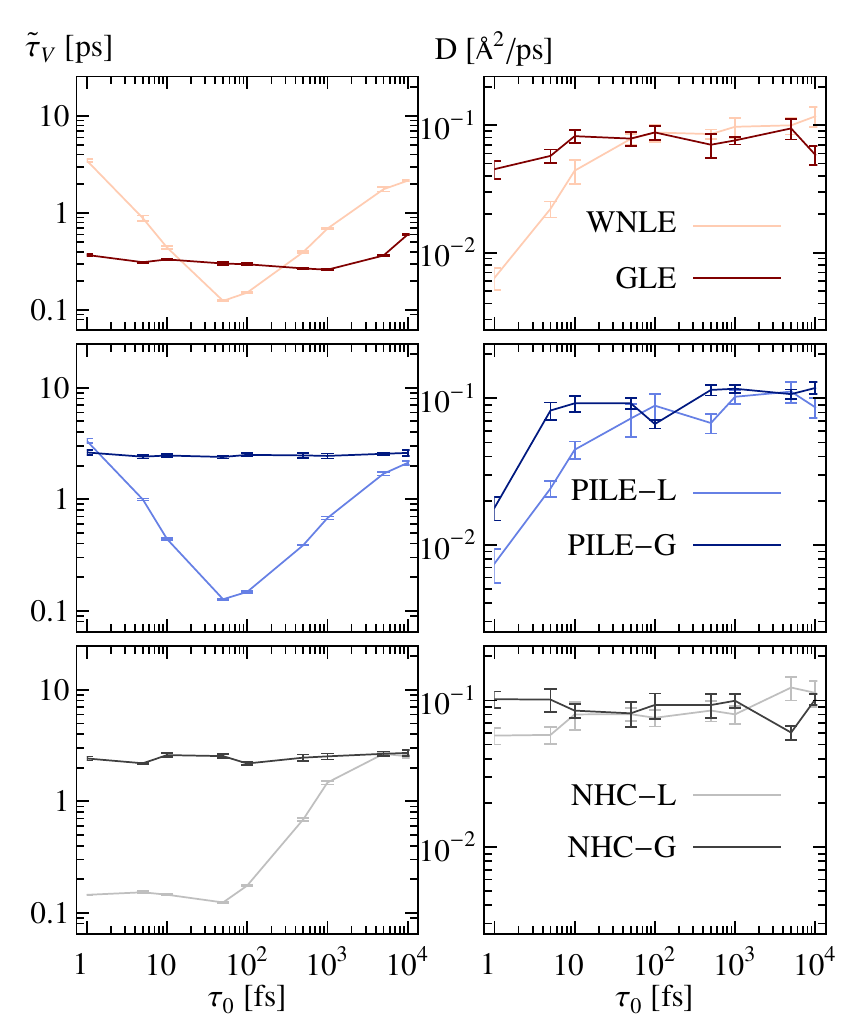}
\caption{
Integral of the absolute value of the normalized autocorrelation function $\tilde{\tau}$ (Eq.~64) for the potential energy of a hydrogen atom interstitial in a palladium lattice (left) and the diffusion coefficient of the hydrogen atom (right) obtained from path integral molecular dynamics simulations. The six panels show the dependence of these quantities on the thermostat relaxation time $\tau_0$ for the WNLE and GLE thermostats and for local and global versions of the PILE and NHC thermostats.}
\end{figure}

Let us now consider two quantities which are more sensitive to the physical forces on the H atom: its contribution to the potential energy (defined as the energy of the H in Pd system minus the energy of the Pd lattice at the same configuration) and its diffusion coefficient. The correlation time $\tau_V$ of the H atom potential energy and the diffusion coefficient $D$ obtained from the thermostatted PIMD simulations are shown in Fig.~5, where the differences between this problem and liquid water become more apparent. Global schemes are now very inefficient, as the frequency mismatch between the H atom motion and the Pd lattice motion results in very slow   equilibration unless a separate thermostat is coupled to each H atom centroid degree of freedom.  This is clear from the results of the PILE-G and NHC-G thermostats, which do not perform any better for any value of the relaxation time $\tau_0$ than the corresponding local thermostats perform in the limit of weak coupling ($\tau_0=10^4$ fs). We expect that this problem would be exacerbated even further if several interstitial H atoms were present and a global thermostat were to be attached to the total kinetic energy of their centroids, as we did in applying the global schemes to liquid water.

For the local schemes $\tau_V$ responds sharply to changes in the thermostat relaxation time $\tau_0$. This occurs because the range of H atom vibrational frequencies involved is relatively narrow, and by choosing an appropriate friction it is possible to obtain nearly-optimal sampling even with a simple WNLE. The GLE thermostat is almost independent of the choice of $\tau_0$,
at the expense of a slight decrease in efficiency with respect to the optimal white-noise friction. This behavior is consistent with the analytical estimates of the efficiencies of the two thermostats in the harmonic limit (Fig.~1), which show the GLE to have a lower maximum efficiency in exchange for the far broader range of frequencies for which it gives near optimal sampling. 

In contrast to the water calculations we see that the diffusion coefficient of the H interstitial in Pd is less affected by overdamping, which manifests itself in Fig.~5 only when a very small value of $\tau_0$ is used. Unlike water, where diffusion is sensitive to orientational modes which lie at higher frequencies in the spectrum, the diffusion mechanism of H atoms in Pd is a much simpler lattice mediated process. The dynamics can therefore tolerate a higher level of disturbance from the thermostat before being hampered. 

Note finally from Fig.~5 that the deterministic NHC-L thermostat does not suffer from such a dramatic degradation in sampling efficiency for small $\tau_0$ as its stochastic counter-part PILE-L. Indeed, it has previously been observed\cite{Ceriotti10} that for purely harmonic potentials the NHC dynamics yields nearly-optimal efficiency for all frequencies smaller than $\omega_0=1/\sqrt{\beta_n\,Q^{(0)}}=1/2\tau_0$. Since the NHC equations lack rotational invariance, however, their performance depends critically on the relative orientations of the eigenvectors of the Hessian with respect to the directions to which the thermostat is applied.\cite{Ceriotti10} The optimal behavior is only obtained with the correct alignment and the sampling efficiency degrades significantly for other alignments in an anisotropic potential, as we have observed here in the case of liquid water (compare the NHC-L results for $\tilde{\tau}_V$ in Fig.~5 with those for $\tau_V$ in Fig.~2 in the limit of small $\tau_0$). 

\section{Concluding Remarks}

Deterministic NHC thermostats have become the {\em de facto} standard for the difficult problem of canonical sampling in path integral molecular dynamics. Stochastic thermostats are however considerably simpler, and they provide a physically very appealing way to model the interaction of a system with a heat bath. Recent advances in the application of stochastic methods to various molecular dynamics sampling problems have therefore encouraged us to reconsider these methods as an alternative to Nos\'e-Hoover chains for path integral simulations, as we have done in this paper.

Overall, we believe that the results we have obtained are very promising. For two distinctly different physical systems -- room temperature liquid water and a hydrogen atom interstitial in a palladium lattice -- the simple stochastic PILE thermostat that we have introduced performs just as well as the deterministic NHC thermostat for nearly every property we have considered (see Figs 2-5). The only exception we have found is for the potential energy of a H atom in a Pd lattice in the strong coupling limit (small $\tau_0$), where the sampling efficiency of the NHC thermostat degrades more gently (see the bottom left hand panel in Fig.~5). In all of the other cases we have considered there is essentially no difference in terms of sampling efficiency between our stochastic PILE and a deterministic NHC.

The GLE thermostat we have investigated has also been found to perform very well, especially given that (unlike the NHC and PILE thermostats) it does not exploit an analytic knowledge of the free ring polymer normal modes. A particularly nice feature of this thermostat is that it combines a sufficiently strong coupling to the internal modes of the ring polymer to ensure ergodic dynamics with a more gentle perturbation on the low frequency centroid motion that only mildly inhibits the Hamiltonian diffusion. As a result of this feature, the GLE behaves in many respects as a compromise between a global and a local thermostat, and provided its time constant $\tau_0$ is chosen such that the range of fitted frequencies encompasses the entire spectral range of interest  it gives close to optimal sampling in every situation.\cite{Note-on-tau0}

One final comment we should make when assessing the relative merits of these various thermostats concerns their computational cost. The implementation of the Langevin equation is clearly very simple in both its white noise (scalar) and colored noise (matrix) forms, because an exact propagator can be obtained for the linear free-particle LE. In contrast, a multiple time step scheme is mandatory when solving the non-linear Nos\'e-Hoover chain equations if one wants to avoid a drift in the conserved quantity. In an {\em ab initio} PIMD simulation, where the computational effort is dominated by the evaluation of the physical forces acting on each ring polymer bead, the extra effort that is required to solve the Nos\'e-Hoover chain equations will often be negligible. However, in the present simulations the evaluation of the forces was made comparatively inexpensive by the use of empirical force fields and a highly efficient ring polymer contraction scheme.\cite{Markland08b,Markland08c} As a consequence, the different thermostats were found to have quite a significant impact on the computational cost of the calculations.\cite{Note-on-timing}

In summary, we firmly believe that stochastic methods should be reconsidered for use in path integral molecular dynamics (and MD in general), as their sampling efficiencies are comparable to that of the most commonly used deterministic scheme and they have a number of practical advantages. We are certainly now using Langevin equation thermostats in our own PIMD simulations and we expect that the results we have presented in this paper will encourage others to do so also. 


\bibliographystyle{aip}

\begin{thebibliography}{10}
\bibitem{Note-on-Fourier-transforms}
The number of Fourier transforms per time step can be reduced from four to
  three per degree of freedom in the absence of a thermostat by keeping the
  momenta in the normal mode representation and transforming the forces into
  this representation for the updates in Eqs.~(21) to (25); this also
  eliminates the need for the Fourier transforms in the PILE thermostat in
  Eqs.~(27) and~(29).
\bibitem{Note-on-tau0}
A suitable choice of $\tau_0$ can be determined automatically for the GLE
  thermostat in advance of the calculation, as follows. The maximum internal
  frequency of the free ring polymer is at $\omega_{\rm max}=2/(\beta_n\hbar)$.
  Since the GLE we have used has been fit to a frequency range of 4 orders of
  magnitude (see Fig.~1), this implies that we should take $\omega_0$ in
  Eq.~(41) to be $\omega_0=10^{-2}\omega_{\rm max}=0.02/(\beta_n\hbar)$, giving
  $\tau_0=1/(2\omega_0)=25\beta_n\hbar$. For liquid water at 298 K with $n=32$
  ring polymer beads, this gives $\tau_0=20$ fs, and for a hydrogen atom in
  palladium at 350 K with 10 beads it gives $\tau_0=55$ fs; inspection of
  Figs.~2 to~5 shows that these values of $\tau_0$ do indeed give nearly
  optimal sampling for the GLE in every case.
\end{thebibliography}

\end{document}